\documentclass[aps,twocolumn,prx,showpacks,preprintnumber]{revtex4-1}
\usepackage[dvips]{graphicx}
\usepackage{dcolumn}
\usepackage{amsmath,amssymb}
\usepackage{txfonts}
\usepackage{bm}
\usepackage{color}

\begin{document}
\draft

\title{Perturbation analysis on large band gap bowing of dilute nitride semiconductors}

\author{Masato Morifuji$^1$ and Fumitaro Ishikawa$^2$}

\affiliation{$^{1}$Graduate School of Engineering, Osaka University, 2-1 Yamada-oka, 
Suita, Osaka 565-0871, Japan\\
$^{2}$Graduate School of Science and Engineering, Ehime University, 3 Bunkyo-cho, 
Matsuyama, Ehime 790-8577, Japan}

\date{\today}

\begin{abstract}
Contrary to the conventional empirical law, 
band gap of dilute nitride semiconductors decreases with nitrogen concentration.
 In spite of a number of investigations, origin of  this   \lq\lq large band gap bowing",
is still under debate.
In order to elucidate this phenomenon, we investigate change of band edge energies 
of GaN$_x$As$_{1-x}$ due to nitrogen by using the perturbation theory.
It is found that energy shift of conduction band edge is arising from 
mixing between $\Gamma$- and L-states and/or $\Gamma$- and X-states
 induced by displacement of Ga atoms around N.
We also found that the valence band edge state shows upward shift in spite of negative
potential of nitrogen.
These results are well understood from character of the wavefunctions and symmetry of 
the perturbation potential.
\end{abstract}

\maketitle

\section{Introduction}

III-V compound semiconductors containing  small amount of nitrogen atoms 
are attracting much attention.\cite{Kondow,Weyers,Bi,Skierbiszewski,Shan1,Tan,Uesugi}
Extensive studies have been carried out since these materials are promising candidate  
for novel optical devices such as infrared lasers and high efficiency 
solar cells.\cite{Tansu,Tansu2,Wiemer}
In addition to the prospect for device applications, their properties which 
are largely different from other 
semiconductors, also evoke much interest.\cite{Noguchi,Sumiya,Fukushima}
In particular, large band gap bowing is a problem still under debate.
Usually, band gap of a mixed compound is well described as a linear interpolation 
of band gaps of the constituent materials.
Deviation from this empirical law, which is called band gap bowing, 
is sometimes observed but not so large.
However, contrary to the empirical law, 
band gap of some compounds such as GaNAs,GaInNAs and GaNP 
containing small amount of N becomes smaller with the N concentration.
There have been various models and theories to explain this 
phenomenon.\cite{Zhao,Shan2,Wu}
Analysis on the basis of tight-binding theory\cite{Lindsay,OReilly},
empirical pseudopotential method \cite{Bellaiche,Bellaiche2,Mader,Kent} 
and first principle calculations\cite{Timoshevskii,Tan2} have been carried out. 
However, its mechanism  has not been clarified, yet.

In this paper, to elucidate the mechanism of the large band gap bowing,
we carry out analysis on  behavior of the band edges 
for a typical dilute nitride semiconductor  GaN$_x$As$_{1-x}$ (GaNAs).
For this purpose, we carry out perturbation calculations
 regarding effects due to nitrogen as the perturbation 
 along with bulk GaAs wavefunctions as bases.

In the next section, first, we show the model used in this study.
After evaluating matrix elements of the perturbation potential, 
reduction of band gap is calculated. 
Then, by making qualitative interpretation in terms of symmetry of 
the perturbation potential and character of the wavefunctions, 
we show how the reduction of band gap due to nitrogen occurs.

\section{Perturbation analysis}
\subsection{Model}

We  consider an $N_d \times N_d \times N_d$ supercell of GaAs in which 
one of As atoms is replaced by a N atom.
Note that this supercell consists of $4{N_d}^3$ unit cells of the zinc blende structure,
that is,  nitrogen concentration is given by $x=1/(4{N_d}^3)$.
Introduction of the N atom gives rise to change in crystalline potential.
We consider following three factors: (i) shift of atomic potential from that of As to N,
(ii) displacement of Ga atoms adjacent to the N atom, and (iii) displacement of 
As atoms on the second neighboring sites to the N atom.
Then, the perturbation potential is given by  
\begin{align}
\varDelta V({\bm r})
=\varDelta V^{\rm (i)}({\bm r})+\varDelta V^{\rm (ii)}({\bm r})+\varDelta V^{\rm(iii)}({\bm r}),
\label{eq:1}
\end{align}
with
\begin{subequations}
\begin{align}
\varDelta V^{\rm(i)}({\bm r}) &=\left[V_{\rm N}({\bm r}-{\bm R}_I)-V_{\rm As}({\bm r}-{\bm R}_I)\right], \\[2mm]
\varDelta V^{\rm(ii)}({\bm r})  &=\sum_j\left[V_{\rm Ga}({\bm r}+{\bm \tau}-{\bm R}_j-{\bm \xi}_j)
               -V_{\rm Ga}({\bm r}+{\bm \tau}-{\bm R}_j) \right], \\
\textrm{and}\notag \\[2mm]
\varDelta V^{\rm(iii)}({\bm r}) &=\sum_{j'}\left[V_{\rm As}({\bm r}-{\bm R}_{j'}-{\bm \eta}_{j'}),
                -V_{\rm As}({\bm r}-{\bm R}_{j'})\right].
\end{align}
\end{subequations}
where the superscripts (i)-(iii) correspond to the three factors noted above.
That is, $\varDelta V^{\rm(i)}({\bm r})$
 denotes the potential shift from that of As to N, and 
$\varDelta V^{\rm(ii)}({\bm r})$ and $\varDelta V^{\rm(iii)}({\bm r})$
 stem from displacement of atoms adjacent to the  nitrogen.
In these equations, $V_{\rm N}$, $V_{\rm As}$, and $V_{\rm Ga}$ 
are atomic potentials of N, As, and Ga, respectively.
For the atomic potentials, we used empirical pseudopotentials presented 
in  Refs.~\cite{Bellaiche,Bellaiche2}
The vector ${\bm R}_I$ $({\bm R}_j)$ indicates location of a zinc blende unit cell, and
$\bm \tau=(a/4,a/4,a/4)$ ($a$ is the lattice constant) is a vector that
 indicates position of a Ga atom within a unti cell.
Displacement of atoms adjacent to the  nitrogen are written as 
$\bm\xi_j$ and $\bm\eta_{j'}$ for the first neighboring 
Ga and the second neighboring As, respectively.
The indices $j$ and $j'$ run through so that ${\bm R}_j-{\bm \tau}+{\bm \xi}_j$ and 
${\bm R}_{j'}+{\bm \eta}_{j'}$ indicate 
the positions of the first neighboring four Ga atoms 
and the second neighboring twelve As atoms, respectively.
We set ${\bm \xi}_j$ so that the Ga atoms approach toward the N atom by 0.38 \AA.
Similarly, ${\bm \eta}_{j'}$ was determined so that the second neighboring 
As atoms approach toward the N by 0.1 \AA.
These values  were obtained from total energies evaluated 
by the first principle calculations using the CASTEP package.\cite{CASTEP}

\begin{figure}[h]
\includegraphics[width=7.0cm]{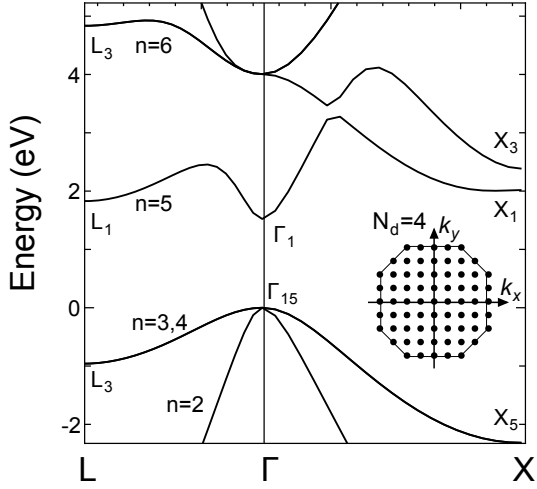}
\caption{Energy band of GaAs calculated by using empirical pseudopotentials.
Inset: Reciprocal lattice vectors to be connected with one another by the perturbation 
potential of the present model for $4\times4\times4$ supercell.
Note that equivalent points are excluded in the following calculations 
though they are plotted in the figures.}
\label{fig:1}
\end{figure}

\subsection{Perturbation matrices and band gap reduction}

In Figure~\ref{fig:1}, we show energy band of GaAs calculated using the empirical 
pseudopotentials along with plane wave basis functions.
Using the wavefunctions of bulk GaAs obtained so far, 
we calculate a matrix element defined by 
\begin{align}
\varDelta V_{n\bm k, n'\bm k'}\equiv \langle\psi_{n \bm k}|\varDelta V|\psi_{n' \bm k'}\rangle,
\label{eq:3}
\end{align}
where $\psi_{n \bm k}({\bm r})$ is a wavefunction of bulk GaAs with a wavevector $\bm k$. 
The index $n$ denotes the band as shown in Fig.~\ref{fig:1} 
where  the highest valence band and the lowest conduction band
are labeled $n=3,4$ and $n=5$, respectively.

In the inset of Fig.~\ref{fig:1}, we show  
$\bm k$-states that are connected with each other by the perturbation potential 
plotted on the first Brillouin zone of the zinc blende structure for the case of $N_d=4$.
These $\bm k$-states discretely distribute as 
\begin{align}
{\bm k}=\frac{2\pi}{N_da}(n_x,n_y,n_z),
\label{eq:4}
\end{align}
with  $n_x, n_y$, and $n_z$ integers between $-N_d$ and $N_d$.
This can be proven as follows:
Since the perturbation potential has translational symmetry with a period $N_{d}a$ 
in all the $x$-, $y$-, and $z$-directions, 
the matrix element $\langle \psi_{n\bm k}|\varDelta V|\psi_{n'\bm k'}\rangle$ 
must be fixed when  $\varDelta V({\bm r})$ is replaced by 
$\varDelta V({\bm r+ \bm R})$ with a lattice vector of the supercell $\bm R$.
From this and the Bloch's theorem, 
the wavevectors $\bm k$ and $\bm k'$ in eq.~(\ref{eq:3}) must satisfy 
a relation ${\bm k}-{\bm k'}=(2\pi)/(N_da)(n_x,n_y,n_z)$.
In the present case, since $\bm k'$ is $\Gamma$, $\bm k$ is given by eq.~(\ref{eq:4}).
Note that some points on the border are equivalent.
For example, $(2\pi/a)(1,0,0)$ and $(2\pi/a)(-1,0,0)$ are identical. 
Hence, one of them must be excluded to avoid double counting,
though both are plotted in the figure.
Excluding such equivalent points, we have $4{N_d}^3$ $\bm k$-points 
in the first Brillouin zone to be mixed due to the perturbation potential;  
 there are 256 points for $N_d=4$ and 2048 points  for $N_d=8$ required for the calculations.
These $\bm k$-points are the points that are folded onto the $\Gamma$-point of the 
Brillouin zone of the $N_d \times N_d \times N_d$ supercell.

\begin{figure}[h]
\includegraphics[width=7.0cm]{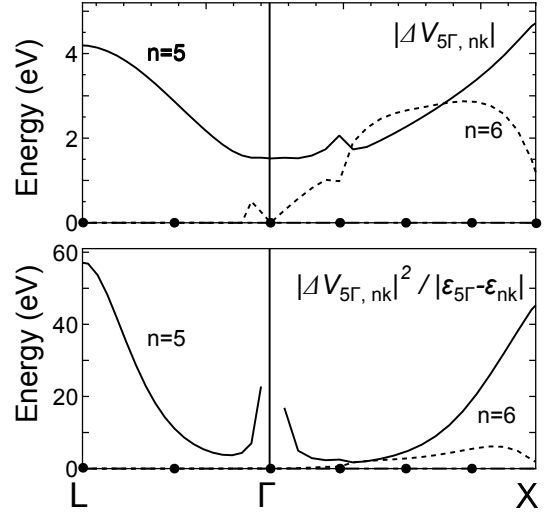}
\caption{Upper panel: Absolute values of perturbation matrix elements 
between the conduction band edge state ($5\Gamma$) and band states 
are plotted as functions of wave vector $\bm k$ along the L-$\Gamma$-X line.
Lower panel: Square of matrix elements divided by energy separation are plotted 
along the L-$\Gamma$-X line.
The dots on the $x$-axis denote $\bm k$-states relevant to the calculations for $N_d=4$.}
\label{fig:2}
\end{figure}

\begin{figure}[h]
\includegraphics[width=7.0cm]{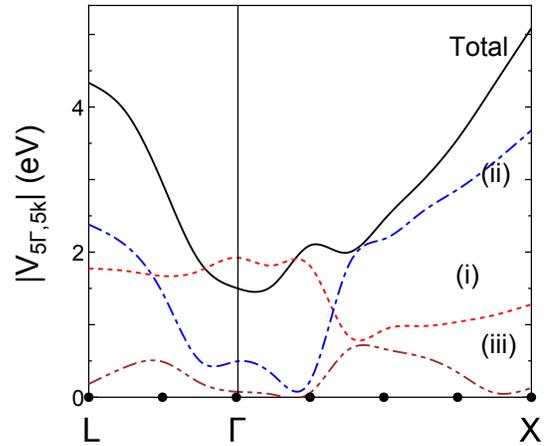}
\caption{Components of the matrix element. The curve labeled (i) --(iii) show matrix elements 
for the factor (i)--(iii) given by eqs.~(2a)--(2c).}
\label{fig:3}
\end{figure}

\begin{figure}[h]
\includegraphics[width=7.0cm]{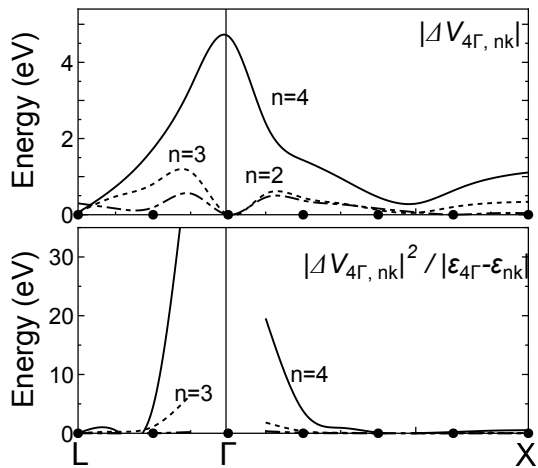}
\caption{Upper panel: Absolute values of perturbation matrix elements 
between the conduction band edge state ($5\Gamma$) and band states 
are plotted as functions of wave vector $\bm k$ along the L-$\Gamma$-X line.
Lower panel: Square of matrix elements over energy separation are plotted 
along the L-$\Gamma$-X line.}
\label{fig:4}
\end{figure}

In the upper panel of Fig.~\ref{fig:2}, we show  $|V_{5 \Gamma, n\bm k}|$, 
i.e., matrix elements between the conduction band edge state 
($\Gamma$-state of the 5th band) and various states.
Note that matrix elements are basicall negative quantities
reflecting deep nitrogen potential, though 
 we plot absolute values since they are complex quantities.
The solid- and dotted- curves show the values 
for the bands $n=5$ and $n=6$, respectively, 
plotted as functions of $\bm k$ along the L-$\Gamma$-X symmetric line of the zinc blende Brillouin zone.
The dots on the horizontal axis show the $\bm k$-states that are relevant to the perturbation calculations,
as we have shown in Fig.~\ref{fig:1}.

In the lower panel, we plot the quantity 
$|V_{5 \Gamma, n\bm k}|^2/|\varepsilon_{5 \Gamma}-\varepsilon_{n\bm k}|$ by 
the solid- and dotted- curves for the band $n=5$ and $n=6$, respectively.
Within the viewpoint of the perturbation theory, magnitude of this quantity corresponds to
energy change  due to the perturbation. 
These curves show that contribution from states around the X- and L-points
 of the lowest conduction band is much larger than that from other states.
As shown in the upper panel, the matrix elements for the 
5th and 6th bands have large values in a wide area of the Brillouin zone.
However, except for the X- and L-states of the 5th band, large energy separation reduces 
this quantity.

To examine further the character of the matrix elements  for the lowest conduction band $n=5$,
we show contributions from the factors (i)--(iii) described 
in the previous section in Fig.~\ref{fig:3}.
In this figure,  the curves labeled  (i)-(iii) show components of the matrix elements due to 
the factors (i)-(iii), respectively.
Note that, since absolute values of complex quantities are plotted, a summation of these components 
does not make the total value.
Around the $\Gamma$-point, the factor (i) the N potential takes the largest value of the three.
Whereas near the X-point, the factor (ii) displacement of Ga atoms is the largest.
At the L-point, the factors (i) and (ii) are comparable though the latter is slightly larger.
Contribution from the factor (iii) displacement of second neighboring As atoms is smaller 
than others in all the region.
As we have shown in Fig.~\ref{fig:2}, contributions from the X- and L-states 
are much larger than that from other state.
This means that the effect due to the displacement of adjacent Ga atom is larger 
than that due to nitrogen potential.
We will discuss this point later in  terms of symmetry of wavefunctions.

In Fig.~\ref{fig:4}, we plot the perturbation matrix elements for the valence states.
Similar to Fig.~\ref{fig:2}, 
$|V_{4 \Gamma, n\bm k}|$ and
$|V_{4 \Gamma, n\bm k}|^2/|\varepsilon_{4 \Gamma}-\varepsilon_{n\bm k}|$
are plotted in the upper and lower panels, respectively.
We note that the degeneracy of the valence states causes a complex situation
because  any linear combinations between the degenerated states satisfy the Schr\"odinger equation.
This gives rise to ambiguity of numerical values.
To avoid such a situation, we made linear combinations between the degenerated states 
so that the states become proper bases of the irreducible representation of the group they belong.
In other words, we set the states to have proper symmetry.
In the calculations, we set $\psi_{4\Gamma}$ to have $x$-symmetry.
Similarly, we set $\psi_{4\bm k}$, $\psi_{3\bm k}$ and $\psi_{2\bm k}$
to have $x$-, $y$-, and $z$-symmetry, respectively, in drawing Fig.~\ref{fig:4}.
%

Different from the conduction states, contribution from states in vicinity of the $\Gamma$-point is large.
Thus, the quantity $|V_{4 \Gamma, 4\bm k}|^2/|\varepsilon_{4 \Gamma}-\varepsilon_{4\bm k}|$
could be very large when $\bm k$ is close to the $\Gamma$-point.
This is because of small denominator arising from heavy mass of the hole states.
However, the states in vicinity of the $\Gamma$ are irrelevant to the energy change unless 
N concentration is very low.
As indicated on the horizontal axis for the case of $N_d=4$, 
the states that are relevant to energy change distribute discretely.
Therefore, the states close to the $\Gamma$ do not affect the energy shift of the band edge state.
When nitrogen concentration is very low, we have to take these states 
into account because when $N_d$ is large the $\bm k$-states denoted by the dots become denser.
However, in this case, as we denote below, the factor $1/(4{N_d}^3)$ added to take nitrogen concentration
into account reduces the contribution from the states close to the $\Gamma$.
%

\begin{figure}[h]
\includegraphics[width=7.0cm]{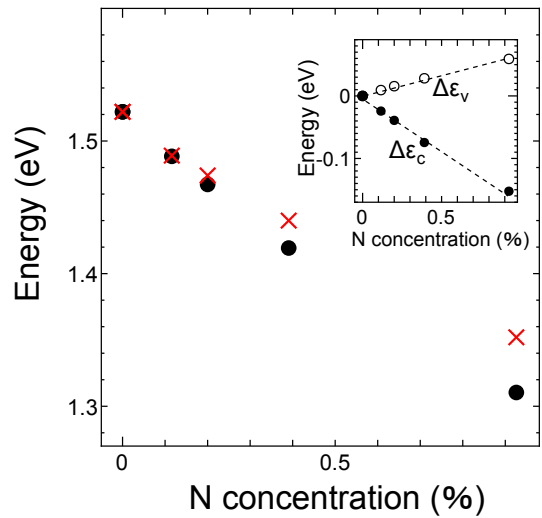}
\caption{Filled circles show calculated band gap as a function of nitrogen concentration..
For comparison, theoretical data from Ref.~[6] 
are also plotted by crosses.
Inset shows energy shifts of the conduction and valence band edge states.}
\label{fig:5}
\end{figure}


Using the perturbation matrix elements, we evaluate energy shift of band edges.
For the conduction band edge, 
the simple perturbation expansion series converges slowly.
This is because, as shown in Fig.~\ref{fig:2}, contribution from the X- and L-states 
is larger than that from the $\Gamma$-state.
This means that higher order terms is larger than the first order term of the perturbation series.
Therefore, we evaluated the band edge energy by diagonalizing 
the Hamiltonian matrix given by 
\begin{align}
{\cal H}_{n\bm k, n'\bm k'} \equiv 
\varepsilon_{n\bm k}\,\delta_{n,n'}\,\delta_{\bm k, \bm k'} + \frac{\varDelta V_{n\bm k, n'\bm k'}}{4{N_d}^3}.
\label{eq:5}
\end{align}
We fixed as  $n=n'=5$ because interband effect is negligible as we have shown in Fig.~\ref{fig:2}.
The factor $1/(4{N_d}^3)$ is necessary to treat N concentration properly.
As we have noted, since the $N_d \times N_d \times N_d$ supercell contains 
$4{N_d}^3$ of zinc blende unit cells, the N concentration is given by $x=1/(4{N_d}^3)$.

For the valence band edge, 
we evaluated energy shift using the secnond order formula
\begin{align}
\varDelta E_v = \frac{1}{4{N_d}^3}V_{4\Gamma, 4\Gamma} 
+ \frac{1}{(4{N_d}^3)^2}\sum_{n=2,3,4}\sum_{\bm k}
\frac{|V_{4\Gamma, n\bm k}|^2}{\varepsilon_{4\Gamma}-\varepsilon_{n\bm k}}.
\label{eq:6}
\end{align}
As we have seen in Fig.~\ref{fig:4}, 
 perturbation matrix elements for the valence states are not so large as those for the conduction states.
As a result, the second order formula yields reasonable results close to the value obtained by 
diagonalizing  perturbation matrix.

In Fig. \ref{fig:5}, we plot calculated band gap by filled circles as a function of nitrogen concentration.
The crosses denote results of first principle calculation extracted from Fig.~6 of Ref.\cite{Tan}.
Reasonable agreement between results of present theory and the first principle calculation 
indicates that the perturbation analysis is valid for dilute nitrides as long as 
N concentration is smaller than 1 \%.
In the inset, we show energy shift of the conduction band edge and valence band edge by filled and open 
circles, respectively. 
In addition to the lowering of the conduction band edge, upward shift of the valence band edge state
is observed.
This seems peculiar since the perturbation potential due to nitrogen is negative.

\begin{figure}[h]
\includegraphics[width=8.0cm]{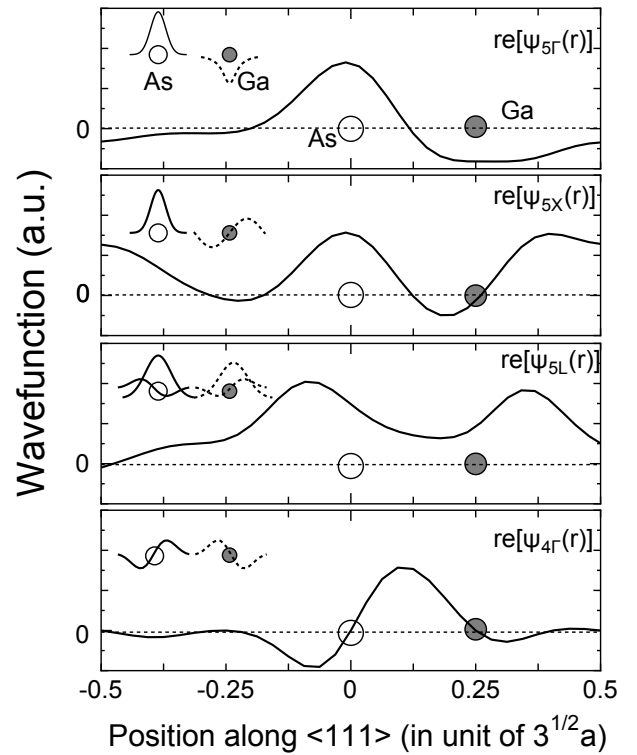}
\caption{Wavefunctions of GaAs. the $\Gamma$, X, and L points are plotted along 
the $\langle 111\rangle$ direction.
Schematic illustrations  to show the character of the wavefunction are also shown in the insets.
The arrows show positions of As (N) and Ga.
The horizontal dotted lines show zero for each wavefunction.}
\label{fig:6}
\end{figure}

\begin{figure}[h]
\includegraphics[width=8.0cm]{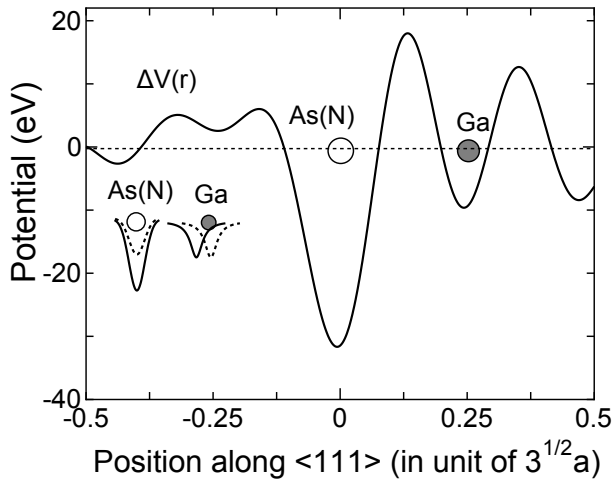}
\caption{Perturbation potential along the  $\langle 111\rangle$ direction 
is plotted by solid curve.
The arrows show positions of As (N) and Ga.
The horizontal dotted line shows zero.
Schematic illustration of potentials with and without nitrogen is also drawn
 to show character of the perturbation.}
\label{fig:7}
\end{figure}

\subsection{Character of wavefunctions and matrix elements}

In Fig.~\ref{fig:6}, we show wavefunctions of GaAs at some symmetry points. 
From the top to bottom, wavefunctions at the $\Gamma$-,
 X-, and L- points of the lowest conduction band and 
the wavefunction of the valence band edge are plotted 
along the $\langle 111 \rangle$ direction.
The As atom locates at the position $0$, and  
 a Ga atom without displacement locates at $+0.25$ as indicated by open and shaded circle, respectively.
For each curve, we also attached schematic illustration to show character of the  wavefunction,
where local wavefunctions of As and Ga are plotted by solid- and dotted- curves, respectively.
The wavefunction of the $\Gamma$ state in the conduction band, $\psi_{5\Gamma}(\bm r)$, 
has peaks at the atom positions, indicating $s$-like character around an atom.
This wavefunction consists of anti-bonding coupling between 
an $s$-like orbital of As and an $s$-like orbital of Ga. 
We also see that wavefunction around As is localized whereas that around Ga is somewhat extended.
In the similar way, we see that 
the wavefunction of the X-state, $\psi_{5X}(\bm r)$, 
consists of anti-bonding coupling between an $s$-like orbital of As and 
a $p$-like orbital of Ga that has a node at the Ga position.
The L-state is somewhat complicated. 
We see that the wavefunction $\psi_{5L}(\bm r)$ has peaks 
at the positions deviated from the atom positions.
This means that, as schematically shown in the inset, the $s$- and $p$-orbitals are mixed 
within an atom.
The large amplitude in the middle of the atoms indicates that this wavefunction has 
bonding character.
The wavefunction $\psi_{4\Gamma}(\bm r)$ has nodes at the atom position, 
meaning $p$-like character around an atom.
We observe that this wavefunction consists of bonding coupling between 
$p$-like orbitals of As and Ga.

In Fig.~\ref{fig:7}, we also plot the perturbation potential along the $\langle 111\rangle$ line.
For each panel, we added schematic atomic potentials with and without nitrogen by dotted- 
and solid-curves, respectively, to illustrate schematically how the perturbation potential arises.
Since the N potential is deeper than As potential, the perturbation potential at the As (N) site is 
negative. 
Due to the spherical character of the pseudopotential, the perturbation potential
 has approximately spherical $s$-like symmetry.
On the other hand, the perturbation potential around Ga, that arises from displacement of Ga atoms,
 is more complicated.
As illustrated in the inset, the displacement of Ga gives rise to potential increase on one side of the 
Ga position and  decrease on another side.
Therefore, the perturbation potential is nearly anti-symmetric around the Ga position, that is, $p$-like 
symmetry along the $\langle 111\rangle$ direction, 
although the potential curve seems complicated due to complexity of pseudopotential 
that is not a simple function.

Comparison between the wavefunctions shown in Fig.~\ref{fig:5} and 
the perturbation potential in Fig.~\ref{fig:6} 
enables us to make interpretation for
 the perturbation matrix elements shown in Fig.~\ref{fig:3}--\ref{fig:5}. 

First, we see the matrix element at the $\Gamma$ state of the conduction band, 
$\varDelta V_{5\Gamma,5\Gamma}$
As shown in Fig.~\ref{fig:3}, contribution of the Nitrogen potential 
is dominant to this first order term $\varDelta V_{5\Gamma,5\Gamma}$.
This is because the wavefunction  has large amplitude at the As (N) position.
The fact that displacement of Ga has small effect to $\varDelta V_{5\Gamma,5\Gamma}$ is well understood;
 As we have noted, potential around Ga has $p$-like symmetry and wavefunction is of $s$-like symmetry.
Therefore integrating $|\psi_{5\Gamma}|^2 \times \Delta V^{\rm (ii)}$, which has
 an odd-like character around Ga, results in small value of the matrix element.

Different from the $\Gamma$-state, the matrix element at the X-state 
effect from Ga is the largest as we have shown in Fig.~\ref{fig:3}. 
This is also understood that the wavefunction of the X-state around the 
Ga site is of $p$-like symmetry.
For the coupling between the $\Gamma$- and X-states $\varDelta V_{5\Gamma,5X}$,
shift of Ga atoms is essential.
$\psi_{5\Gamma}$ has $s$-like symmetry around the Ga atom, 
whereas both $\psi_{5X}$ and $\varDelta V^{\rm (ii)}$have $p$-like symmetry as seen 
in Fig.~\ref{fig:4}. 
Therefore, we foresee that multiplication of these three quantities 
becomes an even function around Ga, which 
enlarges the matrix element $\varDelta V_{5\Gamma,5X}$.

As we have noted, the wavefunction $\psi_{5L}$ has character intermediate between the $\Gamma$- and X-
states; $\psi_{5L}$ has both $s$- and $p$-characters.
This is also well corresponding to the result that $\varDelta V_{5\Gamma,5L}$ arises from both the factor 
(i) and (ii).

We noted that contribution from position shift of the second neighboring As atoms is small. 
This is also explained from symmetry.
Although displacement of As atoms is about 1/3 of that of Ga atoms, 
contribution might be large because of larger number (i.e., 12) of neighboring As atoms.
As we have noted, atom's position change leads to perturbation potential with $p$-like 
symmetry. 
As seen from Fig.~\ref{fig:4}, 
both the $\Gamma$-state and X-state have $s$-like charge distribution around As atoms.
From a simple consideration on symmetry, 
we see that the perturbation matrix element for shift of As 
 with $\bm k$ and $\bm k'$ $\Gamma$ or X has a small value.
{These results indicate that mixing between $\Gamma$ and X or 
between $\Gamma$ and L 
induced by lattice distortion around N is the main factor of the band gap reduction of GaNAs,
rather than the nitrogen potential.}

As for the valence state,
as shown in Fig.~\ref{fig:5}, the valence edge state shows upward shift.
This is innegligible,
since the energy change is as large as 1/3 of that of the conduction state, 
though the valence state is seldom referred.
It seems strange that the valence state shifts upward since the perturbation potential 
is strongly negative at the N position.
As shown in the figure, the valence state has $p$-like symmetry, 
its amplitude is small at the position where the perturbation potential is negative.
The wavefunction has large amplitude at the peripheral where the perturbation is positive.
This makes the first order term positive.
In addition, the second order term of eq.~(\ref{eq:6}) pushes up the valence top state.
%
%
%

\section{Conclusion}

In order to clarify the mechanism of the large band gap bowing of dilute nitride semiconductors, 
we carried out analysis based on the perturbation theory.
It was found that reduction of the conduction band occurs due to intervalley mixing between the $\Gamma$ state
and the X- and L- states.
The intervalley mixing is induced by displacement of Ga atoms around the nitrogen, rather 
than nitrogen potential itself.
This is reasonably understood considering symmetry of wavefunctions.
The displacement of Ga causes potential with $p$-like symmetry,
which causes strong mixing between the $\Gamma$ state and X- (L-) state, since 
the former has $s$-like symmetry and the latter has $p$-like symmetry at the Ga position.
We also found innegligible upward shift of the valence top states.
Although the upward shift in the negative perturbation potential of nitrogen 
seems peculiar, this is also understood from the character of the wavefunction.

%

\end{document}